\documentclass[aps,pra,twocolumn,superscriptaddress,showpacs]{revtex4}

\usepackage{graphicx}

\begin{document}

\title{Virtual noiseless amplification and Gaussian post-selection \\ in continuous-variable quantum key distribution}

\author{Jarom\'{\i}r Fiur\'{a}\v{s}ek}
\affiliation{Department of Optics, Palack\'y University, 17.~listopadu 12, 77146 Olomouc, Czech Republic}

\author{Nicolas J. Cerf}
\affiliation{QuIC, Ecole Polytechnique de Bruxelles, Universit\'e Libre de Bruxelles, 1050 Brussels, Belgium}

\begin{abstract}
The noiseless amplification or attenuation are two heralded filtering  
operations that enable respectively to increase or decrease the mean  
field of any quantum state of light with no added noise, at the cost  
of a small success probability. We show that inserting such  
noiseless operations in a transmission line  
improves the performance of continuous-variable quantum key  
distribution over this line. Remarkably, these noiseless operations  
do not need to be physically implemented but can simply be simulated  
in the data post-processing stage. Hence, \emph{virtual}  noiseless amplification 
or attenuation amounts to perform a Gaussian post-selection, 
which enhances the secure range or tolerable excess noise while keeping the benefits of Gaussian security proofs.
\end{abstract}

\maketitle

Continuous-variable quantum key distribution (CV QKD) based on Gaussian states and homodyne or heterodyne detection can achieve very high secret key rates, see e.g. \cite{Scarani09} for a review. Moreover, its practical implementation does not require single-photon detectors and the system can be made compatible with telecom optical networks \cite{Lodewyck07}. However, although theory predicts that a secure key can be generated for a pure loss channel over an arbitrary large distance \cite{Grosshans03}, the practical range of CV QKD is currently limited to several tens of kilometers by noise and imperfect classical data processing \cite{Fossier09,Jouguet12}.  

In contrast to classical optical networks, losses in quantum communication channels cannot be compensated by usual phase-insensitive amplifiers as the latter inevitably add noise \cite{Caves82}, making the channel insecure. Recently, however, the concept of heralded noiseless quantum amplification has emerged as a novel tool \cite{Ralph08}, which enables one to probabilistically increase the amplitude of a coherent state without adding any extra noise, $|\alpha\rangle \rightarrow |g\alpha\rangle$ with gain $g>1$.  Of course, a natural question arises whether this noiseless amplifier may improve the performance of QKD, especially enhance its secure range. In ref.~\cite{Gisin10}, it was indeed argued that (a double version of) the noiseless amplifier can be beneficial for device-independent quantum cryptography with single photons.

Here, we investigate this question in more general terms. We start from the observation that any physical realization of the noiseless amplifier turns out to be very demanding. Even the proof-of-principle experimental noiseless amplification of weak coherent states requires state-of-the art technology, such as single-photon addition and subtraction, or an auxiliary source of single photons and multiphoton interference \cite{Xiang10,Ferreyrol10,Usuga10,Zavatta11,Osorio12}.  Moreover, the actual success rate of these experiments  is much lower than the theoretical predictions due to various experimental limitations, and the transformation can furthermore only be implemented approximately. Such an approach seems rather impractical in the context of CV QKD where the system should be reasonably simple and robust to allow for field deployment.
 
In this paper, we show that the physical implementation of the noiseless amplifier can be substituted with a suitable data processing, so that the amplification is 
performed only virtually. Just like \emph{virtual} entanglement is used to analyze the security of prepare-and-measure protocols \cite{Grosshans03b}, it appears that \emph{virtual} noiseless amplification may simulate the associated quantum filter and be beneficial to CV QKD. We also turn our attention to a dual quantum filter called noiseless attenuation, which is analogous to noiseless amplification but with a gain lower than one \cite{Micuda}. It probabilistically transforms $|\alpha\rangle \rightarrow |\nu\alpha\rangle$ with gain $\nu<1$, so it is akin to a beam splitter but it effects the same decrease of the mean amplitude for any state with no noise. Noiseless attenuation can also be faithfully emulated by classical postprocessing of the experimental data with a reasonable overhead (as we shall see, noiseless amplification can in principle be emulated arbitrarily well, but an exact emulation is in contrast only possible in the limit of a low success probability due to heavy rejection).

We shall demonstrate that virtual noiseless amplification or attenuation can extend the range of CV QKD over noisy channels. A simple picture, which provides a good intuition of this effect though it is not rigorous, is as follows. The emitter (Alice) preprocesses her signal states by noiselessly attenuating them, thereby making them strongly indistinguishable to an eavesdropper (Eve). At the other end of the line, the receiver (Bob) revives the signal states by noiselessly amplifying them. Since the two quantum filters are Gaussian (trace-decreasing) operations, the pre- and post-selected data appear as if they emerged from a deterministic Gaussian protocol; hence, the security proofs based on the optimality of Gaussian attacks hold \cite{Grosshans04,Navascues06,Patron06}. Somehow, Eve cannot bias the pre- and post-selection filters, and the above ``compaction" of the signal states in the channel can only be detrimental to her. In practice, the pre-selection is not needed (it amounts to reducing the modulation variance), while the post-selection associated with noiseless amplification can be applied virtually on the experimental data.

\begin{figure}[!t!]
\centerline{\includegraphics[width=0.99\linewidth]{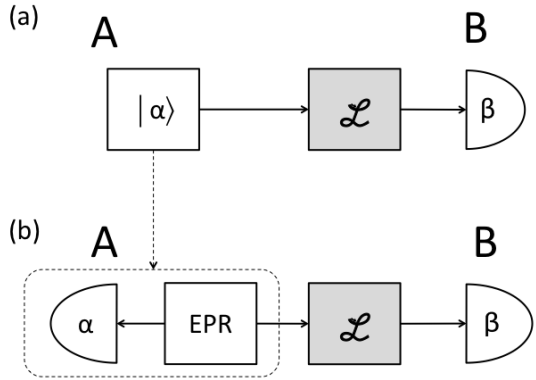}}
\caption{(a) Prepare-and-measure CV QKD protocol with coherent states and heterodyne detection. (b) Equivalent virtual entanglement-based protocol, with heterodyne detection on both sides.}
\end{figure}

\textit{CV QKD protocols}.---Gaussian protocols, to which we restrict here, are based on the Gaussian modulation of Gaussian (coherent or squeezed) states of light and Gaussian (homodyne or heterodyne)  measurements, which gives four possibilities. In the first two protocols, Bob performs homodyne detection, measuring at random the $x$ or $p$ quadrature, while Alice emits a Gaussian-modulated coherent \cite{Grosshans02} or squeezed \cite{Cerf01} state. In the next two, Bob performs heterodyne detection, measuring the $x$ and $p$ quadratures simultaneously, while Alice emits again a  coherent \cite{Weedbrook04} or squeezed \cite{Garcia09} state. Note the existence of a fifth protocol, where Alice sends (mixed) thermal states instead of pure states \cite{Weddbrook10}.

In what follows, we focus on the most symmetric protocol \cite{Weedbrook04}, where Alice emits coherent states $|\alpha\rangle$ and Bob projects onto coherent states $|\beta\rangle$ (heterodyne detection), as illustrated in Fig.~1(a). Alice draws a complex amplitude $\alpha$ from a bivariate Gaussian distribution of variance $V$, and sends $|\alpha\rangle$ to Bob through a quantum channel $\mathcal{L}$ which is controlled by Eve. Then, Bob makes a projective measurement onto coherent states and obtains the outcome $\beta$. After $N$ repetitions of these steps, Alice and Bob  
extract a secret key from the accumulated classical data. From Eve's point of view, this prepare-and-measure protocol is indistinguishable from an entanglement-based  scheme where Alice prepares an entangled two-mode squeezed vacuum state
\begin{equation}
|\Psi_{\mathrm{EPR}}\rangle=\sqrt{1-\lambda^2}\sum_{n=0}^\infty\lambda^{n}|n,n\rangle
\label{EPR}
\end{equation}
with $\lambda^2=2V/(2V+1)$, and performs heterodyne measurement on one mode, see Fig.~1(b).

This virtual entanglement picture \cite{Grosshans03b} is very useful for analyzing the security and understanding the benefit of noiseless amplification. Suppose that $\mathcal{L}$ is a pure loss channel with transmittance $T$. As shown in ref.~\cite{Ralph11}, an entangled state (\ref{EPR}) can be faithfully distributed over $\mathcal{L}$ if Alice sends one mode of a weakly entangled state ($\lambda \ll 1$) to Bob, who noiselessly amplifies his mode. In the considered CV QKD protocol, this would correspond to weak modulation on Alice's side ($V \ll 1$) combined with noiseless amplification on Bob's side, see Fig.~2(a).

\begin{figure}[!t!]
\centerline{\includegraphics[width=0.99\linewidth]{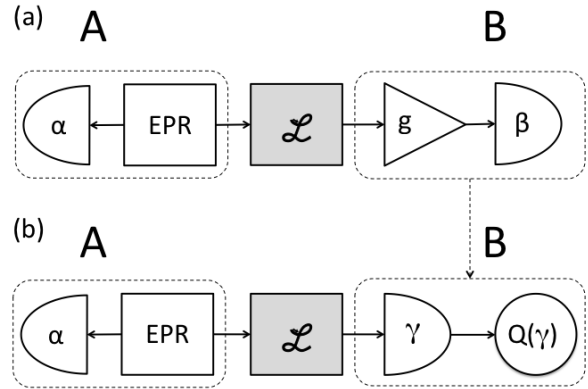}}
\caption{(a) CV QKD with coherent states and heterodyne detection augmented with noiseless amplification of the received signal. (b) Equivalent protocol where noiseless amplification is emulated by post-processing Bob's measurement data.}
\end{figure}

\textit{Virtual noiseless amplification}.---The noiseless amplifier is described by the non-unitary operator $g^{\hat{n}}$, where $\hat{n}$ denotes the photon number operator. Although it is probabilistic, this filter is Gaussian in the sense that it converts a Gaussian state between Alice and Bob in the virtual entanglement picture into another Gaussian state. Forgetting about the modulation variance $V$, this can be viewed as effectively converting the Gaussian channel $\mathcal{L}$ into another Gaussian channel with presumably higher associated performances. Moreover, we can recycle all security proofs and corresponding secret key rates that have be obtained using the Gaussian formalism \cite{Grosshans04,Navascues06,Patron06}.

Unfortunately, $g^{\hat{n}}$ is an unbounded operator for $g>1$, so it cannot be implemented exactly, and, furthermore, its optical implementation is very challenging. Remarkably, these obstacles can be overcome by emulating the noiseless amplifier, which is possible as it is immediately followed by heterodyne measurement. Note that we can consider the noiseless amplifier $g^{\hat{n}}$ at the output of  channel $\mathcal{L}$ to be part of the detection process, see Fig.~2(a). 
Denoting by $\hat{\rho}$ the mixed state at the output of $\mathcal{L}$, Bob obtains (after amplification) the measurement outcome $\beta$ with relative probability
\begin{equation}
P_g(\beta)=\frac{1}{\pi} \; \langle \beta| g^{\hat{n}}  \hat{\rho} \, g^{\hat{n}}|\beta \rangle.
\end{equation}
Using the identity $g^{\hat{n}}|\beta\rangle=e^{(g^2-1)|\beta|^2/2} |g\beta\rangle$, we can write
\begin{equation}
P_g(\beta)=\frac{1}{\pi} \; e^{(g^2-1)|\beta|^2}\langle g\beta | \hat{\rho} |g\beta \rangle.
\label{Pbg}
\end{equation}
If Bob directly measures $\hat{\rho}$ without prior amplification, he gets the outcome $\gamma$  with probability $P(\gamma)=\frac{1}{\pi}\langle \gamma|\hat{\rho}|\gamma\rangle$. By comparing this probability with Eq.~(\ref{Pbg}),
we conclude that Bob can emulate the noiseless amplificater by properly rescaling each measurement outcome $\gamma$ as  $\beta=\gamma/g$, while assigning to it a relative weight  $Q(\gamma)=e^{(1-g^{-2})|\gamma|^2}$, see Fig.~2(b).

This relative weight can be simulated by post-selection, accepting each data $\gamma$ with a probability $P_{\mathrm{acc}}(\gamma)$ that is proportional to $Q(\gamma)$. A difficulty arises here because $Q(\gamma)$ diverges for large $|\gamma|$, which translates the impossibility of implementing a perfect noiseless amplifier. If Alice's modulation $V$ is weak enough, $P(\gamma)$ could be sufficiently narrow so that  $\lim_{|\gamma|\rightarrow \infty }P(\gamma)Q(\gamma)=0$. Then, for a finite number $N$ of data points $\gamma_j$, one can accept each one with probability
\begin{equation}
P_{\mathrm{acc}}(\gamma)=e^{(1-g^{-2})(|\gamma|^2-|\gamma_M|^2)}\leq 1.
\end{equation}
where $|\gamma_M|=\max_j|\gamma_j|$. As derived in the Appendix, the number of accepted data points $N_{\mathrm{acc}}$ grows sublinearly with the size $N$, so unfortunately the rejection rate increases with $N$ and the procedure becomes rather inefficient. Alternatively, one can fix $|\gamma_M|$ independently of $N$. For instance, if $P(\gamma)Q(\gamma)$ is expected to exhibit a distribution with variance $V_{\gamma}$, then one can choose $|\gamma_M|$ as a multiple of $\sqrt{V_{\gamma}}$ (say, 10 standard deviations) and set $P_{\mathrm{acc}}(\gamma)=1$ if $|\gamma| >|\gamma_M|$. 
Assuming a Gaussian distribution of variance $V_B$ for Bob's measurement outcomes $\gamma$, we show in the Appendix that $N_{\mathrm{acc}}$ scales linearly with $N$ in this case, namely
\begin{equation}
\frac{N_{\mathrm{acc}}}{N} \approx \frac{g^2}{g^2+2V_B(1-g^2)} \left[e^{-(1-g^{-2})|\gamma_M|^2}-e^{-\frac{|\gamma_M|^2}{2V_B}}\right]
\end{equation}
Note that this only works if $2V_B < g^2/(g^2-1)$. Given its linear scaling, this second method is more practical than the first one although the data processing does not emulate the exact Gaussian filter because of the finite cut-off, which might complicate the security analysis.

\textit{Virtual noiseless attenuation}.---In view of these difficulties, we also consider a reverse situation where the noiseless amplifier is on Alice's side (replaced, in fact, by a larger $V$) while the noiseless attenuator is on Bob's side (replaced by its virtualization). The noiseless attenuation $\nu^{\hat{n}}$ with $\nu<1$ is a physical operation, which can be implemented by sending the state through a beam splitter of transmittance $\nu^2$ and projecting the auxiliary output port of the beam splitter onto vacuum. Although the efficiency of common single-photon detectors is too low to implement this latter projection with high fidelity, one can faithfully emulate noiseless attenuation with an acceptable overhead. The principle is the same as before. Denoting by $\hat{\rho}$ the state emerging from ${\cal L}$, the relative probability of the measurement outcome $\beta$ after attenuation can be expressed as
\begin{equation}
P_{\nu}(\beta)=\frac{1}{\pi}\langle \beta| \nu^{\hat{n}}  \hat{\rho} \, \nu^{\hat{n}}|\beta \rangle = 
\frac{1}{\pi}e^{-(1-\nu^2)|\beta|^2}\langle \nu\beta | \hat{\rho} |\nu\beta \rangle.
\end{equation}
Since $\nu<1$, we have $e^{-(1-\nu^2)|\beta|^2}<1$, hence no divergence problem. Therefore, 
we can emulate noiseless attenuation by rescaling the measurement outcome $\gamma$ as $\beta=\gamma/\nu$ and accepting the data point with probability $Q(\gamma)=e^{-(\nu^{-2}-1)|\gamma|^2}<1$.  In this way, we post-select a subset of the original data 
that corresponds to a protocol where the signal would be noiselessly attenuated before heterodyne detection. As illustrated in the Appendix, this emulation is efficient in the sense that the number of accepted data points is proportional to the original size, 
\begin{equation}
\frac{N_{\mathrm{acc}}}{N}=\frac{\nu^2}{\nu^2+2V_B(1-\nu^2) },
\end{equation}
The noiseless attenuation is a trace-decreasing Gaussian completely-positive map, so it preserves the Gaussian form of the entangled state between Alice and Bob. Therefore, the security results on deterministic CV QKD protocols with Gaussian modulated coherent states can be applied to this protocol with virtual noiseless attenuation inserted on Bob's side (without the heavy-rejection problem of virtual noiseless amplification).

\begin{figure}[!t!]
\centerline{\includegraphics[width=0.8\linewidth]{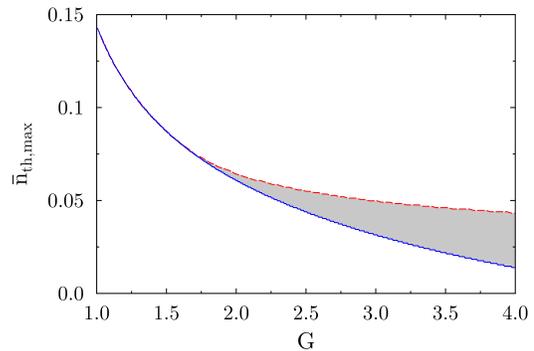}}
\caption{CV QKD over an amplifying channel with gain $G$ and excess thermal noise $\bar{n}_{\mathrm{th}}$. The maximum allowed noise $\bar{n}_{\mathrm{th,max}}$ decreases for increasing $G$. A secret key can be generated if $\bar{n}_{\mathrm{th}}<\bar{n}_{\mathrm{th,max}}$, shown with the blue solid line (standard protocol) or red dashed line (protocol augmented with virtual noiseless attenuation). The grey area indicates the class of channels for which no key can be generated without virtual noiseless attenuation. We optimize over Alice's modulation variance $V$ and Bob's attenuation $\nu$, and assume $\eta=0.9$.}
\end{figure}

\begin{figure}[!t!]
\centerline{\includegraphics[width=0.8\linewidth]{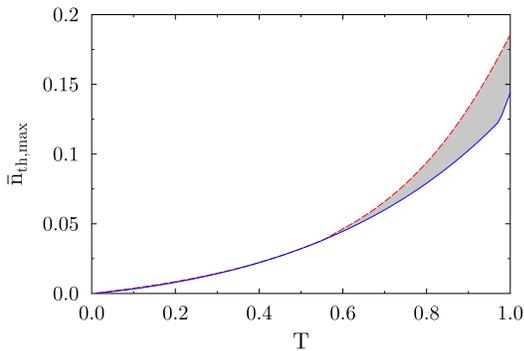}}
\caption{CV QKD over a lossy channel with transmittance $T$ and excess thermal noise $\bar{n}_{\mathrm{th}}$. The maximum allowed noise $\bar{n}_{\mathrm{th,max}}$ decreases for decreasing $T$. A secret key can be generated if $\bar{n}_{\mathrm{th}}<\bar{n}_{\mathrm{th,max}}$, shown with the blue solid line (standard protocol) or red dashed line (protocol augmented with virtual noiseless amplification). The grey area indicates the class of channels for which no key can be generated without virtual noiseless amplification. We optimize over Alice's modulation variance $V$ and Bob's amplification $g$, and assume $\eta=0.9$.}
\end{figure}

\textit{CV QKD with Gaussian post-selection}.---Exploiting that the (Gaussian) quantum filter effected by the noiseless amplifier or attenuator can be emulated during the post-processing stage, we now investigate the benefit of the resulting Gaussian post-selection for CV QKD. As an instructive example, we first consider virtual noiseless attenuation at the output of a Gaussian amplifying channel with excess noise  \cite{Filip08}, see Fig.~3. In the Heisenberg picture, this channel is described by a linear transformation of the annihilation and creation operators,
\begin{equation}
\hat{a}_{\mathrm{out}}=\sqrt{G}\hat{a}_{\mathrm{in}}+\sqrt{G-1}\, \hat{c}^{\dagger},
\end{equation}
where $\hat{a}$ and $\hat{c}$ denote the annihilation operators of the signal and ancilla modes, respectively, and $G$ is the channel gain. This channel is not quantum-noise limited, which is modeled by assuming that the ancilla mode is initially prepared in a thermal state with mean photon number  $\langle c^\dagger c\rangle=\bar{n}_{\mathrm{th}}/(G-1)$, where $\bar{n}_{\mathrm{th}}$ is the mean number of excess thermal photons injected into the signal mode. Sending one part of the entangled state (\ref{EPR}) through this channel yields a mixed two-mode Gaussian state with covariance matrix
\begin{equation}
\gamma_{AB}=\left(
\begin{array}{cc}
a \, I & c \, \sigma_z \\
c \, \sigma_z & b \, I
\end{array}
\right),
\end{equation}
where $a=\cosh(2r)$, $b=G\cosh(2r)+G-1+2\bar{n}_{\mathrm{th}}$, $c=\sqrt{G} \sinh(2r)$, and $r=\tanh^{-1}(\lambda)$. Here, $I$ stands for the $2\times 2$ identity matrix and $\sigma_z$ stands for the third Pauli matrix. The covariance matrix of the Gaussian state obtained conditionally on the success of  $\nu^{\hat{n}}$ (or $g^{\hat{n}}$) can be conveniently calculated by exploring a connection between covariance matrix elements and density matrix elements in Fock basis \cite{Eisert04}, see Appendix. The secret key rate against collective attacks is calculated according to
\begin{equation}
K=\max(\eta I_{AB}-\chi_{AE} , \, \eta I_{AB}-\chi_{BE}),
\end{equation}
where the first (second) term corresponds to direct (reverse) reconciliation, so we choose the protocol that yields the higher secret key rate ($\eta$ is the reconciliation efficiency). Here, $I_{AB}$ is Shannon mutual information between Alice and Bob, while $\chi_{AE}$ ($\chi_{BE}$) is the Holevo quantity between Alice and Eve (Bob and Eve). All these quantities  can be calculated using standard methods (see Appendix) since we know the post-selected virtual Gaussian entangled state shared by Alice and Bob. In Fig.~3, we show the dependence of the maximum allowed thermal noise $\bar{n}_{\mathrm{th,max}}$ on the channel gain $G$. It appears that for sufficiently high $G$, the inclusion of virtual noiseless attenuation is advantageous as a secret key can be generated for a higher level of thermal noise.

We have also performed similar calculations for a lossy channel with transmittance $T$ and excess noise $\bar{n}_{\mathrm{th}}$, see Appendix. Unexpectedly, noiseless attenuation also helps here if $T$ is not too small, although the effect is tiny. We do not plot it here as the effect is similar but much more pronounced by inserting noiseless amplification instead of attenuation on Bob's side. As shown in Fig.~4, the protocol tolerates more thermal noise $\bar{n}_{\mathrm{th}}$ for a fixed $T$ when it is augmented with virtual noiseless amplification. Note that we have neglected the slight non-Gaussianity induced by the post-selection cutoff when calculating the secret key rates. We have also assumed a realistic value $\eta=0.9$ for the efficiency of the classical data reconciliation \cite{Fossier09,Jouguet12}. The advantage of virtual noiseless amplification would be even stronger  for $\eta\to 1$, but this is a rather unrealistic limit as error correction is a highly challenging task in practical CV QKD \cite{Lodewyck07}.

\textit{Conclusion}.---We have demonstrated the improved performance (enhanced secure range or tolerable excess noise) of a CV QKD protocol with coherent states, heterodyne detection, and {\it virtual} noiseless amplification or attenuation. The latter quantum filters do not need to be physically implemented, which would be experimentally quite challenging, but may be simulated by classical post-processing (Gaussian post-selection) of the measured data, making this proposal immediately applicable. Furthermore, since the above quantum filters are Gaussian, the post-selected data can be treated as emerging from an effective Gaussian protocol and the security proofs based on Gaussian extremality still hold.

One may also consider virtual operations in protocols where Bob performs homodyne detection. A noiseless attenuation followed by the projection onto squeezed displaced states can be interpreted as a projection onto squeezed displaced state with lower squeezing and re-scaled displacements. Thus, to simulate noiseless attenuation, we would need to change the detection scheme so that it performs projections onto finitely squeezed states. This could be achieved by employing an eight-port homodyne detection  with unbalanced central beam splitter, such that the amplitude and  
phase quadratures of the state are measured with different precision.  
In view of all this, we anticipate that Gaussian post-selection could be a tool of practical importance in CV QKD.

 \textit{Note:} In the course of completion of this work, the importance of noiseless amplification in CV QKD has been independently demonstrated in  
 \cite{Blandino12} and \cite{Ralph-private}.

\acknowledgments
J.F. acknowledges support from the Czech Science Foundation (P205/12/0577). N.J.C. acknowledges support from the F.R.S.-FNRS under project HIPERCOM.

\newpage

\appendix*
\section {}

\subsection{Gaussian post-selection equivalent to the noiseless amplification}
The heralded noiseless amplifier corresponds to an (unbounded) filtration operator $g^{\hat{n}}$ in Fock space. If it is immediately followed by heterodyne measurement, we have shown that it can be replaced by an appropriate post-processing of the measured data. One must rescale each measurement outcome $\gamma$ as $\beta=\gamma/g$ and assign a relative weight  $Q(\gamma)=e^{(1-g^{-2})|\gamma|^2}$ to it. The unboundedness of the quantum filter translates into the fact that $Q(\gamma)$ diverges for large $|\gamma|$, which makes it difficult to devise a good post-selection procedure. A first possibility is to accept each data point $\gamma$ with probability $P_{\mathrm{acc}}(\gamma)=e^{(1-g^{-2})(|\gamma|^2-|\gamma_M|^2)}\leq 1$, where $|\gamma_M|$ is the maximum value of $|\gamma|$ of the set of size $N$. Then, numerical simulations suggest that assuming Gaussian distribution for Bob's measurement outcomes 
\begin{equation}
P(\gamma)=\frac{1}{2\pi V_B}e^{-\frac{|\gamma|^2}{2V_B}},
\label{Pgm}
\end{equation}
the number of accepted data points $N_{\mathrm{acc}}$ grows with $N$. However, the scaling is only sublinear, $N_{\mathrm{acc}}=N^\kappa$ where $\kappa < 1$, see Fig.~5. This implies $\lim _{N\rightarrow \infty} N_{\mathrm{acc}}/N=0$ and the procedure becomes inefficient for large $N$. 

\begin{figure}[!b!]
\centerline{\includegraphics[width=0.99\linewidth]{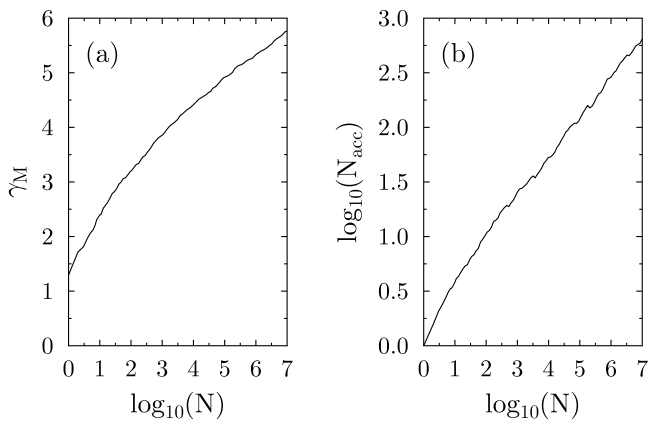}}
\caption{Virtual noiseless amplification. (a) Dependence of Bob's maximum measurement outcome $|\gamma_M|$ on the number of measurements $N$. 
(b) Number of measurement outcomes $N_{\mathrm{acc}}$ preserved after filtration is plotted as a function of $N$. This latter dependence can be approximately described by a power law 
 $N_{\mathrm{acc}} \propto N^{\kappa}$  with $\kappa \approx 0.38$. The graphs were obtained assuming  $V_B=1$ and $g^2=1.5$ and the curves represent averages over $100$ independent 
 simulations of the filtering procedure.}
\end{figure}

A second possibility is to choose a large fixed value for $|\gamma_M|$ and set  $P_{\mathrm{acc}}(\gamma)=1$ if $|\gamma| >|\gamma_M|$. 
The number of data points $N_{\mathrm{acc}}$ preserved after filtering is then lower bounded by
\begin{equation}
N \int_{0}^{2\pi} \int_{0}^{|\gamma_M|} P(\gamma)P_{\mathrm{acc}}(\gamma) |\gamma| \, \mathrm{d}|\gamma|\, \mathrm{d}\phi,
\end{equation}
where $\gamma=|\gamma|e^{i\phi}$. For Gaussian distribution (\ref{Pgm}) one recovers a linear scaling of output data points with $N$,
\[
\frac{N_{\mathrm{acc}}}{N} \geq \frac{g^2}{g^2+2V_B(1-g^2)}\left[e^{-(1-g^{-2})|\gamma_M|^2}-e^{-\frac{|\gamma_M|^2}{2V_B}}\right].
\]
The integral converges and the emulation is possible only if $2(g^2-1)V_B<g^2$.  The emulation of the noiseless amplification is much more practical with this second method, 
but we have to keep in mind that it does not exactly emulate the target Gaussian filtration as the cutoff $|\gamma_M|$ is finite, which might complicate the security analysis.

\subsection{Gaussian post-selection equivalent to the noiseless attenuation}
The idea of replacing the actual noiseless amplifier by some appropriate data post-processing can be applied to the noiseless attenuator too. This amounts to
rescaling the measurement outcome $\gamma$ as $\beta=\gamma/\nu$ and accepting the data point with probability $Q(\gamma)=e^{-(\nu^{-2}-1)|\gamma|^2}$.   The main difference is that there is no divergence problem here since $Q(\gamma)<1$. Therefore, the emulation is efficient in the sense that the number of accepted data points is proportional to the original number of data points, 
\[
\frac{N_{\mathrm{acc}}}{N}=\int_0^{2\pi}\int_0^\infty P(\gamma) e^{-(\nu^{-2}-1)|\gamma|^2} |\gamma|\, \mathrm{d}|\gamma|\, \mathrm{d}\phi.
\]
In particular, for Gaussian distribution of Bob's outcomes (\ref{Pgm}) we obtain
\[
\frac{N_{\mathrm{acc}}}{N}=\frac{\nu^2}{\nu^2+2V_B(1- \nu^2) }.
\]

\subsection{Evaluation of the covariance matrices}
Elements of covariance matrix $\gamma_{AB}$ of a two-mode state $\hat{\rho}_{AB}$ are defined as $\gamma_{AB,jk} =\langle \Delta\hat{R}_{j}\Delta\hat{R}_k + \Delta\hat{R}_{k}\Delta\hat{R}_j\rangle$,
where $\Delta \hat{R}_j=\hat{R}_j-\langle \hat{R}_j\rangle$, $\langle \hat{R}_j\rangle=\mathrm{Tr}[\hat{R}_j\hat{\rho}_{AB}]$ and $\mathbf{\hat{R}}=(\hat{x}_A,\hat{p}_A,\hat{x}_B,\hat{p}_B)$ denotes vector of quadrature operators. Note that variances of quadrature operators of vacuum state are equal to $\frac{1}{2}$, hence the covariance matrix of vacuum is equal to the identity matrix. 

Let us now consider a CV QKD protocol where Alice prepares two-mode squeezed vacuum state 
\begin{equation}
|\Psi_{EPR}\rangle=\sqrt{1-\lambda^2}\sum_{n=0}^\infty \lambda^n|n,n\rangle_{AB}, 
\end{equation}
whose covariance matrix reads
\[
\gamma_{\mathrm{EPR}}=\left(
\begin{array}{cccc}
a & 0 & \sqrt{a^2-1} & 0 \\
0 & a & 0 & -\sqrt{a^2-1} \\
\sqrt{a^2-1} & 0 & a & 0 \\
0 & -\sqrt{a^2-1} & 0 & a 
\end{array}
\right),
\]
and $a=(1+\lambda^2)/(1-\lambda^2)$. 
Alice keeps mode A for heterodyne measurement, while she transmits mode B through the noisy quantum channel $\mathcal{L}$, which can be assumed Gaussian since this is the worst case (optimal collective attack). The covariance matrix changes to 
\begin{equation}
\gamma_{AB}=\left(
\begin{array}{cccc}
a & 0 & c & 0 \\
0 & a & 0 & -c \\
c & 0 & b & 0 \\
0 & -c & 0 & b 
\end{array}
\right),
\end{equation}
where
\begin{equation}
b=Ta+1-T+2\bar{n}_{\mathrm{th}}, \quad c=\sqrt{T(a^2-1)},
\end{equation}
for a lossy channel with transmittance $T\leq 1$ and excess thermal noise $\bar{n}_{\mathrm{th}}$, while
\begin{equation}
b=Ga+G-1+2\bar{n}_{\mathrm{th}}, \quad c=\sqrt{G(a^2-1)},
\end{equation}
for an amplifying channel with gain $G>1$ and excess thermal noise $\bar{n}_{\mathrm{th}}$.

We next determine the covariance matrix of the Gaussian state obtained by applying the quantum filter $g^{\hat{n}}$ to Bob's mode. The cases $g>1$ and $g=\nu < 1$ cover both noiseless amplification and attenuation, respectively. Inspired by \cite{Eisert04}, we exploit the relationship between elements of the covariance matrix $\gamma_{AB}$ and elements of the density matrix $\hat{\rho}_{AB}$ in Fock state basis. The Husimi Q-function of the two-mode state can be expressed as 
\begin{equation}
Q(\mathbf{R})=\frac{\sqrt{\det{\Gamma_{AB}}}}{\pi^2} e^{-\mathbf{R}^T \Gamma_{AB}\mathbf{R}}.
\label{Qfunction}
\end{equation}
where 
\begin{equation}
\Gamma_{AB}=(\gamma_{AB}+I)^{-1},
\label{gGlink}
\end{equation}
 and $I$ denotes the identity matrix. We can write
\begin{equation}
\Gamma_{AB}=\left(
\begin{array}{cccc}
A & 0 & C & 0 \\
0 & A & 0 & -C \\
C & 0 & B & 0 \\
0 & -C & 0 & B 
\end{array}
\right),
\end{equation}
where $A$, $B$, and $C$ are functions of $a$, $b$, $c$. 
Since the Husimi Q-function is a generating function of density matrix elements in Fock basis, one can establish a relationship between elements of matrix $\Gamma_{AB}$ and normalized density matrix elements in Fock basis $\sigma_{jk,lm}=\frac{\rho_{jk,lm}}{\rho_{00,00}}$ \cite{Eisert04}. The action of the filter $\hat{\openone}_{A} \otimes g^{\hat{n}_B}$ is particularly simple in Fock basis,
\begin{equation}
\sigma_{jk,lm}\rightarrow g^{k+m}\sigma_{jk,lm}.
\end{equation}
At the same time, the filtration $g^{\hat{n}}$ is a Gaussian operation which preserves the Gaussian shape of Wigner and Husimi function. 
After a straightforward calculation one finds that the matrix $\Gamma_{AB}$ of the Gaussian state after filtration reads,
\begin{equation}
\Gamma_{AB}^\prime=\left(
\begin{array}{cccc}
A & 0 & gC & 0 \\
0 & A & 0 & -gC \\
gC & 0 & g^2(B-\frac{1}{2})+\frac{1}{2} & 0 \\
0 & -gC & 0 & g^2(B-\frac{1}{2})+\frac{1}{2} 
\end{array}
\right).
\end{equation}
The covariance matrix $\gamma_{AB}^\prime$ of the filtered state can then be obtained by inverting the relation (\ref{gGlink}), 
\begin{equation}
\gamma_{AB}^\prime=(\Gamma_{AB}^{\prime})^{-1}-I.
\end{equation}

\subsection{Evaluation of the secret key rates}

The protocol where Alice prepares coherent states and Bob performs heterodyne detection is formally equivalent to entanglement-based protocol where Alice prepares two-mode squeezed vacuum,
sends one mode to Bob, and both Alice and Bob perform heterodyne detection. In this way, Alice and Bob obtain Gaussian correlated data whose statistics is governed by Husimi Q-function 
(\ref{Qfunction}) and the mutual information between Alice and Bob is given by
\begin{equation}
I_{AB}=\log_2\frac{(a+1)(b+1)}{(a+1)(b+1)-c^2}.
\end{equation}
The Holevo quantity $\chi_{AE}$ can be expressed as
\begin{equation}
\chi_{AE}=S(\hat{\rho}_{AB})-\sum_{j}p_j S(\hat{\rho}_{B,j}),
\label{chiAE}
\end{equation}
where $S(\hat{\rho})$ denotes von-Neumann entropy of quantum state $\hat{\rho}$ and $\hat{\rho}_{B,j}$  denotes density matrix of Bob's mode conditional on $j$th
measurement outcome of Alice. Note that the expression (\ref{chiAE}) follows from the fact that Eve holds purification of the quantum state shared by Alice and Bob hence the 
entropies of Eve's states appearing in the original definition of $\chi_{AE}$ are equal to the entropies of states of Alice and Bob. 
If Alice projects onto coherent  states, then all states $\hat{\rho}_{B,j}$ are Gaussian, have the same covariance matrix
\begin{equation}
\gamma_{B}=\left(
\begin{array}{cc}
b-\frac{c^2}{a+1} & 0   \\
0 & b-\frac{c^2}{a+1} 
\end{array}
\right)
\end{equation}
and differ only in displacements. Therefore all Bob's conditional states have the same entropy $S_B$ and we can write
\begin{equation}
\chi_{AE}=S_{AB}-S_B.
\end{equation}
The von-Neumann entropies of Gaussian states appearing in the above formula can be calculated as follows,
\begin{equation}
S_{AB}=\sum_{j=1}^2 \left(\frac{\mu_j+1}{2}\log_2 \frac{\mu_j+1}{2}-\frac{\mu_j-1}{2}\log_2 \frac{\mu_j-1}{2}\right),
\end{equation}
where $\mu_j$ denote symplectic eigenvalues of the covariance matrix $\gamma_{AB}$, and
\begin{equation}
S_{B}=\frac{\mu+1}{2}\log_2 \frac{\mu+1}{2}-\frac{\mu-1}{2}\log_2 \frac{\mu-1}{2},
\end{equation}
where $\mu=b-c^2/(a+1)$ is the symplectic eigenvalue of covariance matrix $\gamma_B$.

By making suitable replacements in the above formulas we can straightforwardly calculate also $\chi_{BE}$ and the mutual information and Holevo quantity for protocol including noiseless amplification or noiseless attenuation.


\begin{thebibliography}{99}

\bibitem{Scarani09}
V. Scarani, H. Bechmann-Pasquinucci, N. J. Cerf, M. Du\v{s}ek, N. L\"{u}tkenhaus, and M. Peev,
Rev. Mod. Phys. \textbf{81}, 1301 (2009). 

\bibitem{Lodewyck07} J. Lodewyck, M. Bloch, R. Garcia-Patron, S. Fossier, E. Karpov, E. Diamanti, T. Debuisschert, N. J. Cerf, R. Tualle-Brouri, S. McLaughlin, and P. Grangier, Phys. Rev. A \textbf{76}, 042305 (2007).

\bibitem{Grosshans03}
F. Grosshans, G. Van Assche, J. Wenger R. Brouri, N. J. Cerf,  and P. Grangier, Nature \textbf{421}, 238 (2003).

\bibitem{Fossier09} S. Fossier, E. Diamanti, T. Debuisschert, A. Villing, R. Tualle-Brouri, and P. Grangier, New J. Phys. \textbf{11}, 045023 (2009).

\bibitem{Jouguet12}
P. Jouguet, S. Kunz-Jacques, T. Debuisschert, S. Fossier, E. Diamanti, R. Alléaume, R. Tualle-Brouri, P. Grangier, A. Leverrier, P. Pache, and P. Painchault, Opt. Express (in press); arXiv:1201.3744


\bibitem{Caves82}
C. M. Caves, Phys. Rev. D \textbf{26}, 1817-1839 (1982).

\bibitem{Ralph08}
T.C. Ralph and A.P. Lund, in \emph{Quantum Communication Measurement and Computing}, 
Proceedings of 9th International Conference, Ed. A. Lvovsky, 155-160 (AIP, New York 2009); arXiv:0809.0326.  

\bibitem{Gisin10}
N. Gisin, S. Pironio, and N. Sangouard, Phys. Rev. Lett. \textbf{105}, 070501 (2010).


\bibitem{Xiang10} G. Y. Xiang, T. C. Ralph, A. P. Lund, N. Walk, and G. J. Pryde, Nature Phot. \textbf{4}, 316-319 (2010).
\bibitem{Ferreyrol10} F. Ferreyrol, M. Barbieri, R. Blandino, S. Fossier, R. Tualle-Brouri, and P. Grangier, Phys. Rev. Lett. \textbf{104}, 123603 (2010). 
\bibitem{Usuga10} M. A. Usuga, C. R. Muller, C. Wittmann, P. Marek, R. Filip, C. Marquardt, G. Leuchs, and U. L. Andersen, Nature Phys. \textbf{6}, 767-771 (2010).
\bibitem{Zavatta11} A. Zavatta, J. Fiur\'{a}\v{s}ek, and M. Bellini, Nature Phot. \textbf{5}, 52-56 (2011). 
\bibitem{Osorio12} C. I. Osorio, N. Bruno, N. Sangouard, H. Zbinden, N. Gisin, and R. T. Thew, arXiv:1203.3396.


\bibitem{Grosshans03b}
F. Grosshans, N. J. Cerf, J. Wenger, R. Tualle-Brouri, and P. Grangier, Quantum Inf. Comput. \textbf{3}, 535 (2003).

\bibitem{Micuda} M. Mi\v{c}uda, I. Straka, M. Mikov\'{a}, M. Du\v{s}ek, N. J. Cerf, J. Fiur\'{a}\v{s}ek, and M. Je\v{z}ek, submitted.


\bibitem{Grosshans04} F. Grosshans and N. J. Cerf, Phys. Rev. Lett. \textbf{92}, 047905 (2004).

\bibitem{Navascues06} M. Navascu\'{e}s, F. Grosshans, and A. Acin, Phys. Rev. Lett. \textbf{97}, 190502 (2006).

\bibitem{Patron06} R. Garc\'{\i}a-Patr\'{o}n and N.J. Cerf, Phys. Rev. Lett. \textbf{97}, 190503 (2006). 

\bibitem{Grosshans02} F. Grosshans and P. Grangier,  Phys. Rev. Lett. \textbf{88}, 057902 (2002).

\bibitem{Cerf01} N. J. Cerf, M. Levy, and G. Van Assche, Phys. Rev. A \textbf{63}, 052311 (2001). 

\bibitem{Weedbrook04}
C. Weedbrook, A.M. Lance, W.P. Bowen, T. Symul, T.C. Ralph, and P.K. Lam, Phys. Rev. Lett. \textbf{93}, 170504 (2004);
S. Lorenz, N. Korolkova, and G. Leuchs, Appl. Phys. B \textbf{79}, 273 (2004).

\bibitem{Garcia09} R. Garcia-Patron and N. J. Cerf, Phys. Rev. Lett. \textbf{102}, 130501 (2009).

\bibitem{Weddbrook10} C. Weedbrook, S. Pirandola, S. Lloyd, and T. C. Ralph, Phys. Rev. Lett. \textbf{105}, 110501 (2010).


\bibitem{Ralph11} T. C. Ralph, Phys. Rev. A \textbf{84}, 022339 (2011). 


\bibitem{Filip08}
R. Filip, Phys. Rev. A \textbf{77}, 032347 (2008).

\bibitem{Eisert04}
J. Eisert, D.E. Browne, S. Scheel, M.B. Plenio, Annals of Physics (NY) \textbf{311}, 431 (2004).


\bibitem{Blandino12} R. Blandino, A. Leverrier, M. Barbieri, J. Etesse, P. Grangier, and R. Tualle-Brouri, 9th workshop on continuous-variable quantum information processing, Frederiksdal, April 27-30, 2012; arXiv:1205.0959.

\bibitem{Ralph-private} T. C. Ralph \textit{et al.}, 9th workshop on continuous-variable quantum information processing, Frederiksdal, April 27-30, 2012.



\end{thebibliography}
\end{document}